\documentclass[useAMS,usenatbib]{mn2e}
\usepackage{times}
\usepackage{graphicx}
\usepackage{subfigure}
\usepackage{psfrag}
\usepackage{amsmath}
\usepackage{amssymb}

\def\reff@jnl#1{{\rm#1\/}}
\def\aj{\reff@jnl{AJ}}                 
\def\araa{\reff@jnl{ARA\&A}}           
\def\apj{\reff@jnl{ApJ}}               
\def\apjl{\reff@jnl{ApJ}}              
\def\apjs{\reff@jnl{ApJS}}             
\def\ao{\reff@jnl{Appl.Optics}}        
\def\apss{\reff@jnl{Ap\&SS}}           
\def\aap{\reff@jnl{A\&A}}              
\def\aapr{\reff@jnl{A\&A~Rev.}}        
\def\aaps{\reff@jnl{A\&AS}}            
\def\azh{\reff@jnl{AZh}}               
\def\baas{\reff@jnl{BAAS}}             
\def\jrasc{\reff@jnl{JRASC}}           
\def\memras{\reff@jnl{MmRAS}}          
\def\mnras{\reff@jnl{MNRAS}}           
\def\pra{\reff@jnl{Phys.Rev.A}}        
\def\prb{\reff@jnl{Phys.Rev.B}}        
\def\prc{\reff@jnl{Phys.Rev.C}}        
\def\prd{\reff@jnl{Phys.Rev.D}}        
\def\prl{\reff@jnl{Phys.Rev.Lett}}     
\def\pasp{\reff@jnl{PASP}}             
\def\pasj{\reff@jnl{PASJ}}             
\def\qjras{\reff@jnl{QJRAS}}           
\def\skytel{\reff@jnl{S\&T}}           
\def\solphys{\reff@jnl{Solar~Phys.}}   
\def\sovast{\reff@jnl{Soviet~Ast.}}    
\def\ssr{\reff@jnl{Space~Sci.Rev.}}    
\def\zap{\reff@jnl{ZAp}}               
\def\nat{\reff@jnl{Nature}}            

\title[Two companions orbiting HIP 5158] {Bayesian evidence for two
  companions orbiting HIP 5158} \author[F. Feroz, S.T. Balan and
  M.P. Hobson] {F.~Feroz\thanks{E-mail: f.feroz@mrao.cam.ac.uk},
  S.~T.~Balan and M.~P.~Hobson\\Astrophysics Group, Cavendish
  Laboratory, JJ Thomson Avenue, Cambridge CB3 0HE, UK\\}

\date{Accepted ---. Received ---; in original form \today}
\pagerange{\pageref{firstpage}--\pageref{lastpage}}
\pubyear{2011}

\begin{document}
\label{firstpage}
\maketitle

\begin{abstract}
We present results of a Bayesian analysis of radial velocity (RV) data
for the star HIP 5158, confirming the presence of two companions and
also constraining their orbital parameters. Assuming Keplerian orbits,
the two-companion model is found to be $e^{48}$ times more probable
than the one-planet model, although the orbital parameters of the
second companion are only weakly constrained.  The derived orbital
periods are $345.6 \pm 2.0$ d and $9017.8 \pm 3180.7$ d respectively,
and the corresponding eccentricities are $0.54 \pm 0.04$ and $0.14 \pm
0.10$. The limits on planetary mass $(m \sin i)$ and semimajor axis
are $(1.44 \pm 0.14$ $M_{\rm J}, 0.89 \pm 0.01$ AU) and $(15.04 \pm
10.55$ $M_{\rm J}, 7.70 \pm 1.88$ AU) respectively. Owing to the large
uncertainty on the mass of the second companion, we are unable to
determine whether it is a planet or a brown dwarf. The remaining
`noise' (stellar jitter) unaccounted for by the model is $2.28 \pm
0.31$ m/s. We also analysed a three-companion model, but found it to
be $e^{8}$ times less probable than the two-companion model.
\end{abstract}


\section{Introduction}\label{sec:intro}

Extrasolar planetary research has made great advances in the last decade as a result of the data gathered by several ground and space based telescopes
and so far more than 500 extrasolar planets have been discovered. More and more planets with large orbital periods and small velocity amplitudes are
now being detected due to remarkable improvements in the accuracy of RV measurements. With the flood of new data, more powerful statistical techniques
are being developed and applied to extract as much information as possible. Traditionally, the orbital parameters of the planets and their
uncertainties have been obtained by a two stage process. First the period of the planets is determined by searching for periodicity in the RV data
using the Lomb-Scargle periodogram (\citealt{1976Ap&SS..39..447L, 1982ApJ...263..835S}). Other orbital parameters are then determined using
minimisation algorithms, with the orbital period of the planets fixed to the values determined by Lomb-Scargle periodogram.

Bayesian methods have several advantages over traditional methods, for example when the data do not cover a complete orbital phase of the planet.
Bayesian inference also provides a rigorous way of performing model selection which is required to decide the number of planets favoured by the data.
The main problem in applying such Bayesian model selection techniques is the computational cost involved in calculating the Bayesian evidence.
Nonetheless, Bayesian model selection has the potential to improve the interpretation of existing observational data and possibly detect yet
undiscovered planets. Recent advances in Marko-Chain Monte Carlo (MCMC) techniques (see e.g. \citealt{MacKay}) have made it possible for Bayesian
techniques to be applied to extrasolar planetary searches (see e.g. \citealt{2005ApJ...631.1198G, 2005AJ....129.1706F, 2007ASPC..371..189F,
2009MNRAS.394.1936B}). \citet{2010arXiv1012.5129F} recently presented a new Bayesian method for determining the number of extrasolar planets, as well
as for inferring their orbital parameters, without having to calculate directly the Bayesian evidence for models containing a large number of planets,
although this method is not required for the analysis of the simple system considered here.

HIP 5158 is a K-type main sequence star at a distance of $45 \pm 4$ pc with mass $0.780 \pm 0.021$ $M_{\sun}$
\citep{2010A&A...512A..48L}. In this paper, we present a Bayesian analysis of the existing, high-precision RV
measurements of HIP 5158 from the HARPS instrument \citep{2003Msngr.114...20M}, given in
\citet{2010A&A...512A..48L}, who detected a planet with period $345.72 \pm 5.37$~d. They did not find any
correlation of the RV with the bispector span of the cross correlation function for HIP 5158. They
also did not find any periodicity in the stellar activity indicator $\log R_{\rm HK}^{\prime}$, making it unlikely
for any long period RV variability to be caused by stellar activity. Moreover, they noticed an additional
quadratic drift in the RV data which they inferred to indicate the presence of another body orbiting the star,
but they were not able to constrain its orbital parameters.

The outline of this paper is as follows. We give a brief introduction to Bayesian inference in Sec.~\ref{sec:bayesian} and describe our method
for calculating the number of planets favoured by the data. Our model for calculating RV is described in Sec.~\ref{sec:RV}. In
Sec.~\ref{sec:bayes_RV} we describe our Bayesian analysis methodology. We apply our method to RV data sets on HIP 5158 in
Sec.~\ref{sec:results} and present our conclusions in Sec.~\ref{sec:conclusions}.

\section{Bayesian inference}\label{sec:bayesian}

Our planet finding methodology is built upon the principles of Bayesian inference, and so we begin by giving a
brief summary of this framework. We refer the reader to \citet{2010arXiv1012.5129F} for a more thorough discussion
on Bayesian object detection methods.

Bayesian inference methods provide a consistent approach to the estimation of a set of parameters
$\mathbf{\Theta}$ in a model (or hypothesis) $H$ for the data $\mathbf{D}$. Bayes' theorem states that
\begin{equation} 
\Pr(\mathbf{\Theta}|\mathbf{D}, H) = \frac{\Pr(\mathbf{D}|\,\mathbf{\Theta},H)\Pr(\mathbf{\Theta}|H)}{\Pr(\mathbf{D}|H)},
\label{eq:bayes}
\end{equation}
where $\Pr(\mathbf{\Theta}|\mathbf{D}, H) \equiv P(\mathbf{\Theta})$ is the posterior probability distribution of
the parameters, $\Pr(\mathbf{D}|\mathbf{\Theta}, H) \equiv \mathcal{L}(\mathbf{\Theta})$ is the likelihood,
$\Pr(\mathbf{\Theta}|H) \equiv \pi(\mathbf{\Theta})$ is the prior, and $\Pr(\mathbf{D}|H) \equiv \mathcal{Z}$ is
the Bayesian evidence.

In parameter estimation, the normalising evidence factor can be ignored, since it is independent of the
parameters $\mathbf{\Theta}$, and inferences are obtained by taking samples from the (unnormalised) posterior
using standard MCMC methods, where at equilibrium the chain contains a set of samples from the parameter
space distributed according to the posterior. This posterior constitutes the complete Bayesian inference of the
parameter values, and can be marginalised over each parameter to obtain individual parameter constraints.

In contrast to parameter estimation problems, for model selection the evidence needs to be evaluated and is
simply the factor required to normalize the posterior over $\mathbf{\Theta}$:
\begin{equation}
\mathcal{Z} = \int{\mathcal{L}(\mathbf{\Theta})\pi(\mathbf{\Theta})}d^D\mathbf{\Theta},
\label{eq:Z}
\end{equation} 
where $D$ is the dimensionality of the parameter space. As the average of the likelihood over the prior, the
evidence is larger for a model if more of its parameter space is likely and smaller for a model with large areas
in its parameter space having low likelihood values, even if the likelihood function is very highly peaked. Thus,
the evidence automatically implements Occam's razor: a simpler theory with compact parameter space will have a
larger evidence than a more complicated one, unless the latter is significantly better at explaining the data.
The question of model selection between two models $H_{0}$ and $H_{1}$ can then be decided by comparing their
respective posterior probabilities given the observed data set $\mathbf{D}$, as follows
\begin{equation}
R = \frac{\Pr(H_{1}|\mathbf{D})}{\Pr(H_{0}|\mathbf{D})}
  = \frac{\Pr(\mathbf{D}|H_{1})\Pr(H_{1})}{\Pr(\mathbf{D}| H_{0})\Pr(H_{0})}
  = \frac{\mathcal{Z}_1}{\mathcal{Z}_0} \frac{\Pr(H_{1})}{\Pr(H_{0})},
\label{eq:R}
\end{equation}
where $\Pr(H_{1})/\Pr(H_{0})$ is the a priori probability ratio for the two models, which can often be set to
unity but occasionally requires further consideration. The natural logarithm of the ratio of posterior model
probabilities (sometimes termed the posterior odds ratio) provides a useful guide to what constitutes a
significant difference between two models:
\begin{equation}
\Delta \ln R = \ln \left[ \frac{\Pr(H_{1}|\mathbf{D})}{\Pr(H_{0}|\mathbf{D})}\right]
=\ln \left[ \frac{\mathcal{Z}_1}{\mathcal{Z}_0}\frac{\Pr(H_{1})}{\Pr(H_{0})}\right].
\label{eq:Jeffreys}
\end{equation}

The evaluation of the number of objects favoured by the data is a model selection problem. By considering a {\em
series} of models $H_{N_{\rm obj}}$, each with a {\em fixed} number of objects, i.e. with $N_{\rm
obj}=0,1,2,\ldots$, one can infer $N_{\rm obs}$ by identifying the model with the largest marginal posterior
probability $\Pr(H_{N_{\rm obj}}|\mathbf{D})$. The probability associated with $N_{\rm obj} = 0$ is often called
the `null evidence' and provides a baseline for comparison of different models. Indeed, this approach has been
adopted previously in exoplanet studies (see e.g. \citet{2010MNRAS.403..731G}), albeit using only lower-bound
estimates of the Bayesian evidence for each model.

Evaluation of the multidimensional integral in Eq.~\ref{eq:Z} is a challenging numerical task. Standard
techniques like thermodynamic integration are extremely computationally expensive which makes evidence evaluation
at least an order of magnitude more costly than parameter estimation. The nested sampling approach, introduced by
\citet{skilling04}, is a Monte Carlo method targeted at the efficient calculation of the evidence, but also
produces posterior inferences as a by-product. \citet{feroz08} and \citet{multinest} built on this nested
sampling framework and have recently introduced the {\sc MultiNest} algorithm which is very efficient in sampling
from posteriors that may contain multiple modes and/or large (curving) degeneracies and also calculates the
evidence. This technique has greatly reduces the computational cost of Bayesian parameter estimation and model
selection and has already been applied to several model selections problem in astrophysics (see e.g.
\citealt{2008arXiv0810.0781F, 2009MNRAS.398.2049F, 2009CQGra..26u5003F}). We employ this technique in this paper.

\section{Modelling Radial Velocities}\label{sec:RV}

Observing planets at interstellar distances directly is extremely difficult, since the planets only reflect
the light incident on them from their host star and are consequently many times fainter. Nonetheless, the
gravitational force between the planets and their host star results in the planets and star revolving around
their common centre of mass. This produces doppler shifts in the spectrum of the host star according to its RV,
the velocity along the line-of-sight to the observer. Several such measurements, usually over an extended period
of time, can then be used to detect extrasolar planets.

Following the formalism given in \cite{2009MNRAS.394.1936B}, for $N_{\rm p}$ planets and ignoring the
planet-planet interactions, the RV at an instant $t_{\rm i}$ observed at $j$th observatory can be calculated as:
\begin{equation}
v(t_{\rm i},j) = V_{\rm j} - \sum_{\rm p=1}^{N_{\rm p}} K_{\rm p} \left[\sin(f_{\rm i, p} +
  \varpi_{\rm p}) 
+ e_{\rm p} \sin(\varpi_{\rm p})\right],
\label{eq:RV}
\end{equation}
where
\begin{eqnarray*}
V_{\rm j} & = & \mbox{systematic velocity with reference to $j$th observatory},\\
K_{\rm p} & = & \mbox{velocity semi-amplitude of the $p$th planet},\\
\varpi_{\rm p} & = & \mbox{longitude of periastron of 
the $p$th planet},\\
f_{\rm i, p} & = & \mbox{true anomaly of the $p$th planet},\\
e_{\rm p}& = & \mbox{orbital eccentricity of the $p$th planet},\\ [-1mm]
& & \mbox{start of data taking, at which
periastron occurred.}
\end{eqnarray*}
Note that $f_{\rm i, p}$ is itself a function of $e_{\rm p}$, the
orbital period $P_{\rm p}$ of the $p$th planet, and the fraction
$\chi_{\rm p}$ of an orbit of the $p$th planet, prior to the start of
data taking, at which periastron occurred.  While there is a unique mean
line-of-sight velocity of the center of motion, it is important to
have a different velocity reference $V_{\rm j}$ for each
observatory/spectrograph pair, since the velocities are measured
differentially relative to a reference frame specific to each
observatory.

\begin{table*}
\begin{center}
\begin{tabular}{|c|c|c|c|c|}
\hline
Parameter & Prior & Mathematical Form & Lower Bound & Upper Bound \\ 
\hline\hline
$P$ (days) & Jeffreys & $\frac{1}{P \ln(P_{\rm max}/P_{\rm min})}$ & $0.2$ & $365,000$ \\
$K$ (m/s) & Mod. Jeffreys & $\frac{(K+K_0)^{-1}}{\ln(1+(K_{\rm max}/K_0)(P_{\rm min}/P_{\rm
i})^{1/3}(1/\sqrt{1-e_{\rm i}^2}))}$ & $0$ & $K_{\rm max}(P_{\rm min}/P_{\rm i})^{1/3}(1/\sqrt{1-e_{\rm i}^2})$\\
$V$ (m/s) & Uniform & $\frac{1}{V_{\rm min}-V_{\rm max}}$ & $-K_{\rm max}$ & $K_{\rm max}$ \\
$e$ & Uniform & $1$ & $0$ & $1$ \\
$\varpi$ (rad) & Uniform & $\frac{1}{2\pi}$ & $0$ & $2\pi$ \\
$\chi$ & Uniform & $1$ & $0$ & $1$ \\
$s$ (m/s) & Mod. Jeffreys & $\frac{(s+s_0)^{-1}}{\ln(1+s_{\rm max}/s_0)}$ & $0$ & $K_{\rm max}$ \\ \hline
\end{tabular}
\end{center}
\caption{Prior probability distributions.}
\label{tab:priors}
\end{table*}

We also model the intrinsic stellar variability $s$ (`jitter'), as a source of uncorrelated Gaussian noise in
addition to the measurement uncertainties. Therefore for each planet we have five free parameters: $K$, $\varpi$,
$e$, $P$ and $\chi$. In addition to these parameters there are two nuisance parameters $V$ and $s$, common to all
the planets.

These orbital parameters can then be used along with the stellar mass $m_{\rm s}$ to calculate the length $a$ of
the semi-major axis of the planet's orbit around the centre of mass and the planetary mass $m$ as follows:
\begin{eqnarray}
a_{\rm s}\sin i & = & \frac{K P \sqrt{1-e^2}}{2 \pi}, \\
\label{eq:as}
m \sin i & \approx &\frac{Km_{\rm s}^{\frac{2}{3}} P^{\frac{1}{3}} \sqrt{1-e^2}}
{(2\pi G)^\frac{1}{3}},
\label{eq:mp}\\
a & \approx & \frac{m_{\rm s} a_{\rm s} \sin i}{m\sin i},
\label{eq:a}
\end{eqnarray}
where $a_{\rm s}$ is the semi-major axis of the stellar orbit about the centre-of-mass and $i$ is the angle
between the direction normal to the planet's orbital plane and the observer's line of sight. Since $i$ cannot be
measured with RV data, only a lower bound on the planetary mass $m$ can be estimated.

\section{Bayesian Analysis of Radial Velocity Measurements}\label{sec:bayes_RV}

There are several RV search programmes looking for extrasolar planets. The RV measurements consist of the time
$t_{\rm i}$ of the $i$th observation, the measured RV $v_{\rm i}$ relative to a reference frame and the
corresponding measurement uncertainty $\sigma_{\rm i}$. These RV measurements can be analysed using Bayes'
theorem given in Eq.~\ref{eq:bayes} to obtain the posterior probability distributions of the model parameters
discussed in the previous section. We now describe the form of the likelihood and prior probability
distributions.

\subsection{Likelihood function}\label{sec:RV_like}

As discussed in \cite{2007MNRAS.374.1321G}, the errors on RV measurements can be treated as Gaussian and
therefore the likelihood function can be written as:
\begin{equation}
\mathcal{L}(\Theta) = \prod_{\rm i} \frac{1}{\sqrt{2\pi(\sigma_{\rm i}^2 + s^2)}} \exp\left[
-\frac{(v(\Theta;t_{\rm i}) - v_{\rm i})^2}{2(\sigma_{\rm i}^2 + s^2)} \right],
\label{eq:L}
\end{equation}
where $v_{\rm i}$ and $\sigma_{\rm i}$ are the $i^{\rm th}$ RV measurement and its corresponding uncertainty
respectively, $v(\Theta;t_{\rm i})$ is the predicted RV for the set of parameters $\Theta$, and $s$ is intrinsic
stellar variability. A large value of $s$ can also indicate the presence of additional planets, e.g. if a
two-planet system is analysed with a single-planet model then the velocity variations introduced by the second
planet would act like an additional noise term and therefore contribute to $s$.

\subsection{Choice of priors}\label{sec:RV_priors}

For parameter estimation, priors become largely irrelevant once the data are sufficiently constraining, but for
model selection the prior dependence always remains. Therefore, it is important that priors are selected based on
physical considerations. We follow the choice of priors given in \cite{2007MNRAS.374.1321G}, as shown in
Table~\ref{tab:priors}.

The modified Jeffreys prior,
\begin{equation}
\Pr(\theta|H) = \frac{1}{(\theta + \theta_0) \ln(1+\theta_{\rm max}/\theta_0)},
\label{eq:modjeff}
\end{equation}
behaves like a uniform prior for $\theta \ll \theta_0$ and like a Jeffreys prior (uniform in $\log$) for $\theta
\gg \theta_0$. We set $K_0 = s_0 = 1$ m/s and $K_{\rm max} = 2129$ m/s, which corresponds to a maximum
planet-star mass ratio of $0.01$.

\section{Results}\label{sec:results}

\begin{table}
\begin{center}
\begin{tabular}{|c|r|r|r|}
\hline
$N_{\rm p}$ & $\ln \mathcal{Z}$ & $s$ (m/s) \\ 
\hline\hline
$1$ & $30.05 \pm 0.14$ & $10.31 \pm 1.09$ \\
$2$ & $78.19 \pm 0.15$ & $ 2.28 \pm 0.31$ \\
$3$ & $70.68 \pm 0.16$ & $ 2.28 \pm 0.28$ \\ \hline
\end{tabular}
\end{center}
\caption{The evidence and jitter values for the system HIP 5158. The null evidence ($-255.3$) has been subtracted
from each $\ln \mathcal{Z}$ value.}
\label{tab:Z_HIP5158}
\end{table}

\begin{figure*}
\begin{center}
\includegraphics[width=1.9\columnwidth]{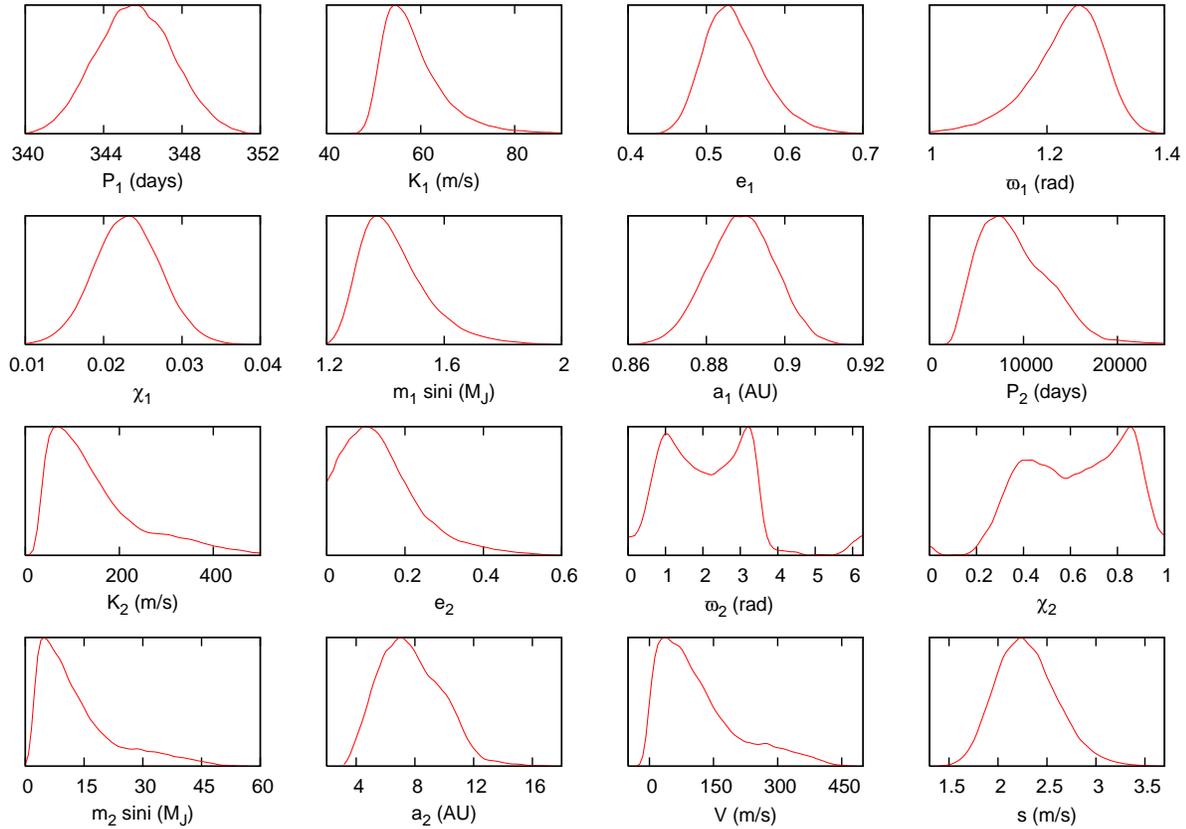}
\caption{1-D marginalised posterior probability distributions for the parameters of the two bodies found
orbiting HIP 5158.}
\label{fig:HIP5158_1D}
\end{center}
\end{figure*}

As discussed in Sec.~\ref{sec:bayesian}, the number of companions orbiting a star can be determined by analysing
the RV data, starting with $N_{\rm p} = 0$ and increasing it until the Bayesian evidence for $N_{\rm p} = n$
companions starts to drop off. The resulting evidence and jitter values for HIP 5158 RV data, after subtracting a
mean RV of $14.9105$ km/s from it, are presented in Table~\ref{tab:Z_HIP5158}. We can clearly see
$N_{\rm p} = 2$ is the strongly favoured model. Adopting the 2-companion model, the estimated parameter values
are listed in Table~\ref{tab:Z_HIP5158_theta} while the 1-D marginalised posterior probability distributions are
shown in Fig.~\ref{fig:HIP5158_1D}. As mentioned in Sec.~\ref{sec:intro}, stellar activity is unlikely to be the
source of any long period RV variability of HIP 5158. Moreover, the inferred period of HIP 5158 c is more
than two orders of magnitude higher than the rotation period of the star ($P_{\rm rot}(\log R_{\rm
HK}^{\prime})=42.3$ d, \citealt{2010A&A...512A..48L}). We therefore conclude that it is unlikely for the second 
signal to be generated by the photospheric activity of HIP 5158.

\begin{figure}
\begin{center}
\includegraphics[width=0.9\columnwidth]{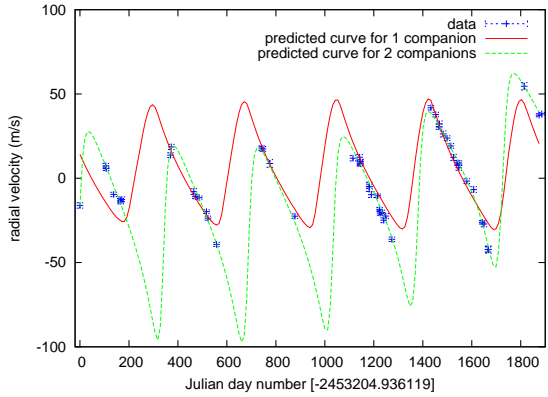}
\caption{RV measurements, with $1\sigma$ errorbars, and the mean fitted RV curve with
1 (red solid line) and 2-companions (green dashed line) for HIP 5158.} 
\label{fig:HIP5158_RV}
\end{center}
\end{figure}

\begin{figure}
\begin{center}
\includegraphics[width=0.9\columnwidth]{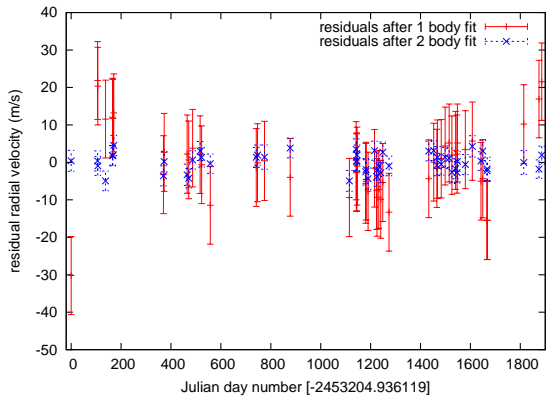}
\caption{RV residuals, with $1\sigma$ errorbars that include the
  estimated stellar jitter, after fitting for one (red solid bars) and
  two companions (blue dotted bars).}
\label{fig:HIP5158_residuals}
\end{center}
\end{figure}

\begin{figure}
\begin{center}
\includegraphics[width=1\columnwidth]{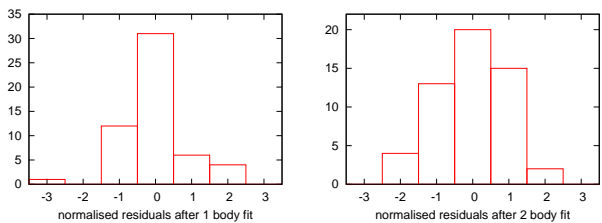}
\caption{Histograms of normalised RV residuals, as defined in Eq.~\ref{eq:norm_residuals}.}
\label{fig:HIP5158_hist}
\end{center}
\end{figure}

The mean RV curve for 1 and 2-companion models are overlaid on the RV
measurements in Fig.~\ref{fig:HIP5158_RV}. The RV residuals after
fitting for 1 and 2 companions along with nuisance parameters
(constant systematic velocity and stellar jitter) are shown in
Fig.~\ref{fig:HIP5158_residuals}, and are obtained by subtracting the
estimated mean RV from the measured RV and adding the estimated
stellar jitter in quadrature to the quoted RV uncertainties. The
1-companion fit is clearly very poor due to the presence of quadratic
drift in the RV data, whereas the residuals from the 2-companion fit 
are consistent with a noise-only model.
Fig.~\ref{fig:HIP5158_hist} shows histograms of normalised data
residuals, defined as follows:
\begin{equation} v_{\rm r, i} = \frac{v_{\rm i} - \bar{v}_{\rm i}}{\sqrt{ \sigma_{\rm i}^2 + \bar{s}^2}},
\label{eq:norm_residuals}
\end{equation}
where $v_{\rm i}$, $\sigma_{\rm i}$ and $\bar{v}_{\rm i}$ are the
measured RV, measured RV uncertainty and the estimated mean RV
respectively, all at time $t_{\rm i}$ and $\bar{s}$ is the estimated
mean stellar jitter. The histogram for the normalised residuals in the
1-companion fit (left panel) has a similar width to the corresponding
histogram for the 2-companion fit (right panel), but it should be
noted that $\bar{s}$ appearing in the normalisation of
Eq.~\ref{eq:norm_residuals} is much smaller in the latter case (see
Table~\ref{tab:Z_HIP5158})

\begin{table}
\begin{center}
\begin{tabular}{|c|r|r|}
\hline
Parameter			& HIP 5158 b		& HIP 5158 c	      \\
\hline\hline
$P$ (days) 			& $345.63 \pm 1.99$	& $9017.76 \pm 3180.74$ \\
	 			&$(341.82, 349.53)$     & $(3567.13, 18193.94)$         \\
	 			&$(344.74)$             & $(11135.63)$         \\
$K$ (m/s) 			& $ 59.05 \pm 7.73$	& $ 170.54 \pm  110.17$ \\
	 			&$(49.89, 77.47)$       & $(36.59, 416.85)$         \\
	 		        &$(56.65)$              & $(218.49)$           \\
$e$ 				& $  0.54 \pm 0.04$	& $   0.14 \pm  0.10$ \\
	 			&$(0.47, 0.63)$         & $(0.01, 0.40)$         \\
	 		        &$(0.54)$               & $(0.00)$            \\
$\varpi$ (rad) 			& $  1.23 \pm 0.07$	& $   2.48 \pm  1.32$ \\
	 			&$(1.06, 1.34)$         & $(0.33, 5.90)$         \\
	 		        &$(1.24)$               & $(2.23)$            \\
$\chi$	 			& $  0.02 \pm 0.00$	& $   0.57 \pm  0.23$ \\
	 			&$(0.02, 0.02)$         & $(0.12, 0.95)$         \\
	 		        &$(0.02)$               & $(0.59)$            \\
$m \sin i$ ($M_{\rm J}$)	& $  1.44 \pm 0.14$	& $  15.04 \pm 10.55$ \\
	 			&$(1.26, 1.77)$         & $(2.34, 41.49)$         \\
	 		        &$(1.34)$               & $(19.53)$            \\
$a$ (AU)			& $  0.89 \pm 0.01$	& $   7.70 \pm  1.88$ \\
	 			&$(0.87, 0.91)$         & $(4.22, 12.50)$         \\
	 		        &$(0.87)$               & $(8.83)$            \\ \hline
\end{tabular}
\end{center}
\caption{Estimated parameter values for the two bodies found orbiting
  HIP 5158. The estimated values in the first row are quoted as $\mu
  \pm \sigma$ where $\mu$ and $\sigma$ are the posterior mean and
  standard deviation respectively. The interval in the second row is
  the $95\%$ confidence interval and the number in parenthesis in the
  third row is the maximum-likelihood parameter value.}
\label{tab:Z_HIP5158_theta}
\end{table}

Comparing our parameter values with those given in \cite{2010A&A...512A..48L}, we see that our parameter
estimates for planet HIP 5158 b are in very good agreement. The large uncertainties on the parameter estimates of
HIP 5158 c are because not enough phase has been covered by the RV measurements. Nevertheless, we can still
determine that the period of the second companion is greater than 10 years with the $95\%$ confidence interval
being $(3567, 18194)$ days. The corresponding $95\%$ confidence intervals for minimum mass and semi-major axis of
the second companion are $(2.3, 41.5)$ $M_{\rm J}$ and $(4.2, 12.5)$ AU respectively and therefore we can not
determine whether it is a planet or a brown dwarf.

\section{Conclusions}\label{sec:conclusions}

We have presented a Bayesian analysis of HIP 5158 RV data. Using Bayesian model selection, we have found a very
strong signal for the presence of two companions orbiting HIP 5158. Our estimated orbital parameters for the
planet HIP 5158 b are in excellent agreement with the values given in \cite{2010A&A...512A..48L}. We determined
the orbital period of HIP 5158 c to be at least 10 years but the presence of large uncertainties in its orbital
parameters do not allow us to determine whether it is a planet or a brown dwarf.

\section*{Acknowledgements}

This work was carried out largely on the {\sc Cosmos} UK National Cosmology Supercomputer at DAMTP, Cambridge and the Darwin Supercomputer of the University
of Cambridge High Performance Computing Service ({\tt http://www.hpc.cam.ac.uk/}), provided by Dell Inc. using Strategic Research Infrastructure Funding from
the  Higher Education Funding Council for England. We would like to thank the anonymous referee for useful comments. FF is supported by a Research Fellowship
from Trinity Hall, Cambridge. STB acknowledges support from the Isaac Newton Studentship.

\bibliographystyle{mn2e}
\bibliography{references}

\appendix

\label{lastpage}

\end{document}